\documentclass{aa}

\usepackage{graphicx}
\usepackage{float}
\usepackage{balance}

\usepackage{txfonts}

\begin{document}

   \title{Non-thermal emission from microquasar/ISM interaction}


   \author{P. Bordas\inst{1}, V. Bosch-Ramon\inst{2}, J.~M. Paredes\inst{1} \and M. Perucho\inst{3} }

\institute{Departament d'Astronomia i Meteorologia and Institut de Ci\`encies del Cosmos (ICC), Universitat de Barcelona (UB/IEEC), Mart\'{i} i Franqu\`{e}s 1, E08028 Barcelona (Spain) \and Max Planck Institut f\"ur Kernphysik, Saupfercheckweg 1, Heidelberg 69117, Germany \and Max-Planck-Institut f\"ur Radioastronomie, Auf dem H\"ugel 69, 53121 Bonn, Germany}


   \date{Received  ??, 2008; accepted ??, 2008}

   \abstract
{The interaction of microquasar jets with their environment can produce non-thermal radiation as is the case for extragalactic outflows impacting on their surroundings. Observational evidences of jet/medium interactions in galactic microquasars have been collected in the past years, although few theoretical work has been done regarding the resulting non-thermal emission.} 
{In this work we investigate the non-thermal emission produced in the interaction between microquasar jets and their environment, and the physical conditions for its production.}
{We have developed an analytical model based on those successfully applied to extragalactic sources. The jet is taken to be a supersonic and mildly relativistic hydrodynamical outflow. We focus on the jet/shocked medium structure when being in its adiabatic phase, and assume that it grows in a self-similar way. We calculate the fluxes and spectra of the radiation produced via synchrotron, Inverse Compton and relativistic Bremsstrahlung processes by electrons accelerated in strong shocks. A hydrodynamical simulation is also performed to further investigate the jet interaction with the environment and check the physical parameters used in the analytical model.} 
{For reasonable values of the magnetic field, and using typical values for the external matter density, the non-thermal particles could produce significant amounts of radiation at different wavelengths, although they do not cool mainly via radiative channels but through adiabatic losses. The physical conditions of the analytical jet/medium interaction model are consistent with those found in the hydrodynamical simulation.}
{Microquasar jet termination regions could be detectable at radio wavelengths for current instruments sensitive to $\sim$~arcminute scales. At X-rays, the expected luminosities are moderate, although the emitter is more compact than the radio one. The source may be detectable by XMM-Newton or Chandra, with 1--10~arcseconds of angular resolution. The radiation at gamma-ray energies may be within the detection limits of the next generation of satellite and ground-based instruments.}


   \maketitle
%

\section{Introduction}

\noindent Microquasars (MQ) are radio emitting X-ray binaries (REXB) that display relativistic ejections (Mirabel \& Rodr\'iguez \cite{Mirabel99}). The binary system is formed by a compact object, a neutron star or a black hole, and a normal (non-degenerate) star. The relativistic ejections may be transient or persistent (Rib\'o \cite{Ribo2005}), with duty cycles of jet activity that change from source to source. The steady jets are formed during the so-called Low/Hard state (Fender et al. \cite{Fender2001}), and are expected to be only mildly relativistic (Gallo et al. \cite{Gallo2003}). Presently, about 15 MQs are known in the Milky Way (Paredes \cite{Paredes2005}), although some authors have proposed that many if not all REXBs could be MQs (e.g. Fender \cite{Fender2004a}). Thus, the real number of MQs in our galaxy could be notably higher.

\noindent MQs are considered scaled-down versions of distant quasars since they present many of the characteristics of these extragalactic sources. They serve as suitable scenarios for understanding a number of processes, such as mass accretion onto the compact object or jet formation and evolution, in timescales inaccessible in the case of quasars. In analogy with radio-loud quasars, which show jets impacting on the intracluster medium, one may expect to find radiative signatures due to strong shocks in the termination regions of MQ jets as well. Hot spots and double-lobe morphologies are common features of the powerful extragalactic FRII sources (Faranoff \& Riley \cite{Faranoff1974}), in which a variety of large-scale non-thermal structures are revealed at radio wavelengths. In the case of MQs, convincing evidence of jet interaction with the interstellar medium (ISM) are not numerous, partially due to the relatively small number of known sources. Nevertheless, hints or evidence of such interactions have been observed, at different wavelengths, in SS~433 (Zealey et~al. \cite{zealey80}), 1E~1740.7$-$2942 (Mirabel et al. \cite{mirabel92}), XTE~J1550$-$564 (Corbel et~al. \cite{corbel02}), Cygnus~X-3 (Heindl et al. \cite{heindl03}), Cygnus~X-1 (Mart\'i et al. \cite{marti96}; Gallo et~al. \cite{gallo05}), GRS~1915+105 (Kaiser et al. \cite{Kaiser04}), H1743$-$322 (Corbel et~al. \cite{corbel05}), LS~I~+61~303 (Paredes et al. \cite{paredes07}a), and Circinus~X-1 (Tudose et al. \cite{tudose06}), although only some theoretical work has been done regarding the hydrodynamics of the interaction, or the resulting non-thermal emission (e.g. Aharonian \& Atoyan \cite{aharonian98}; Vel\'azquez \& Raga \cite{velazquez00}; Heinz \& Sunyaev \cite{heinz02b}; Bosch-Ramon et al. \cite{bosch05}; Perucho \& Bosch-Ramon \cite{Perucho08}).

\noindent MQ ejections can transport large amounts of kinetic energy and momentum very far from the binary system. The age of the source times the jet kinetic power, $\tau_{\rm MQ}\times Q_{\rm jet}$, can be as high as $10^{12}$~s~($3\times 10^{4}$~yr)~$\times~10^{37}-10^{39}$~erg~s$^{-1} \sim 10^{49}-10^{51}$~erg (see. e.g., Cygnus~X-1, Gallo et al. \cite{gallo05}; SS~433, Zealey et al. \cite{zealey80}). Even for low radiative efficiencies converting this energy into non-thermal emission, MQ jet termination regions could produce significant fluxes if they were at distances of few kpc. It could well be that MQs were associated, via interaction with their surrounding medium, to some of the extended non-thermal radio sources detected in the Galaxy of unknown origin (e.g., Paredes et al. \cite{Paredes07}b).

\noindent In this work, we investigate whether a typical MQ fulfills the conditions to be detectable when interacting with the surrounding external gas. We adopt some reasonable assumptions for the MQ power and age, and the density of the ISM, as well as for the non-thermal energy and magnetic field equipartition fraction in the interaction regions. We study the case of a high-mass system in order to see the role of a massive and hot companion. In the case of a low-mass system, the radiation field of the companion star and therefore the Inverse Compton (IC) contribution would be much lower.

\noindent This paper is organized as follows. In Section~2, we present an analytical interaction model to characterize the three shocked zones; for each region, the main properties of the non-thermal emitters are studied, and the physical conditions and geometry are established. In Sect.~3, we calculate the total emission produced through synchrotron, relativistic Bremsstrahlung, and IC processes for the different set of parameters used in the model.  We carried out hydrodynamical simulations to study in more detail the jet evolution when interacting with the ambient medium. These simulations allow us also to test and enforce the validity of the assumptions of the analytical model. This is done in Sect.~4. The details of the hydrodynamical simulations can be found in Appendix A. Finally, in Sect.~5 we discuss the obtained results, giving detectability predictions at different energy bands.


\section{Model description}

\noindent We use a model adapted from those applied to extragalactic sources in order to investigate the interactions of MQ jets with their environment. Falle (\cite{Falle1991}) and Kaiser \& Alexander (\cite{Kaiser97}) have developed an analytical model where self-similarity is used to characterize the evolution of the jet/medium interaction structures in radio galaxies. We have closely followed their work, with minor modifications required for the application of the model to the MQ context.

\noindent We consider twin conical jets emerging from the central source in opposite directions. The ejecta begin to decelerate when the mass of the swept up external gas becomes similar to the mass carried by the jet. Two shocks are formed at the jet tip: a forward shock (the bow shock) propagating into the ISM and a reverse shock directed inwards into the jet material. Matter crossing the reverse shock inflates the cocoon, which protects the jet from disruption due to turbulent gas entrainment. Moreover, a reconfinement shock is also formed in the jet at the point where its pressure equals that of the cocoon.

\noindent In our scenario, all the shocks are assumed to be strong. The density and pressure of the shocked material are calculated using the Rankine-Hugoniot conditions for a strong shock at each front. We note that since the conditions in the shocked regions evolve with time, all related physical variables will also change with time. Therefore, the non-thermal particle population is calculated considering the time dependence of the radiative and adiabatic losses (see Sect.~2.3). Given that all the considered shocks are weakly relativistic, we adopt an adiabatic index $\hat{\gamma} = 5/3$. We assume the presence of a randomly oriented magnetic field $B$ in the downstream regions, derived taking the magnetic energy density to be $\sim$~10~\% ($\eta=0.1$) of the thermal energy density. In each shock region, the fraction of kinetic luminosity assumed to be transferred to non-thermal particles is taken to be 1~\% ($\chi=0.01$).

\subsection{The properties of the three emitting zones} 

\begin{figure}

   \centering 

   \includegraphics[width=8cm]{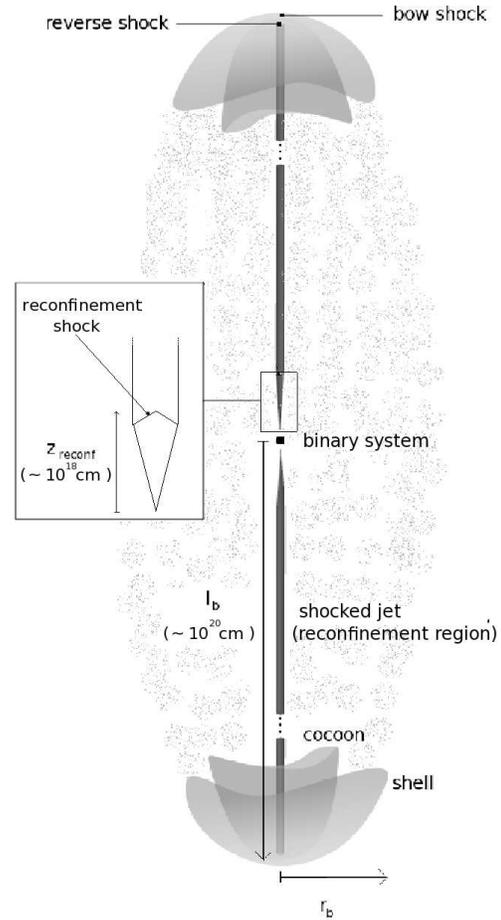}

      \caption{Scheme of the model (not to scale) representing the three different zones. The central binary system is located at the center. Two jets emerge from it and extent outwards until they are effectively decelerated at a distance $\sim 10^{20}$ cm. Jet material that crosses the reverse shock inflates the cocoon, which expands exerting work on the shocked ISM. A contact discontinuity separates the cocoon and the shocked ISM, the latter being further limited by the bow shock.}

         \label{fig1}

   \end{figure}

\noindent We focus our studies in three separated emitting zones. The first one, the shell region, corresponds to the ISM material swept up by the bow shock. The second one, the cocoon region, corresponds to the material of the jet crossing the reverse shock. The third zone accounts for the shocked jet material after crossing the conical reconfinement shock. A sketch of the model is shown in  Fig.~\ref{fig1}.

\subsubsection{\it{Bow shock}}

\noindent   We are interested in the case when the bow shock can accelerate particles, and we concentrate on its adiabatic
phase (see Sedov \cite{Sedov1959}), which implies that the inertia of the medium must be large enough to take a significant fraction of the energy and momentum from the jet. Although in the extragalactic context it is thought that the non-thermal radio emission comes solely from the shocked jet material in the cocoon, it is worth to explore the possibility to have also a contribution from the shocked ISM material. After the interaction structures have reached a characteristic length-scale $l_{\rm 0}=[Q_{\rm jet}^{2}/\rho_{\rm ISM}^{2}c^{6}(\Gamma_{\rm jet}-1)^3]^{1/4}\,$, where $Q_{\rm jet}$ is the jet power, $\rho_{\rm ISM}$ is the ISM mass density, $\Gamma_{\rm jet}$ is the jet bulk Lorentz factor and $c$ is the speed of light, the equations for the length and velocity of the bow shock at a given source age $t_{\rm MQ}$ are (see Falle (\cite{Falle1991}), Kaiser \& Alexander (\cite{Kaiser97})):

\begin{equation}
l_{\rm b}=c_{1}\left(\frac{Q_{\rm jet}}{\rho_{\rm ISM}}\right)^{1/5}t_{\rm MQ}^{3/5}\,, 
\end{equation}
and
\begin{equation}
v_{\rm b}=\frac{d}{dt}(l_{\rm b})=\frac{3l_{\rm b}}{5t_{\rm MQ}}\,.
\end{equation}
 respectively. $c_{1}\approx 1$ is a dimensionless constant that depends on the adiabatic index of the jet material and the geometry of the bow shock. The pressure in the shell can be calculated as $P_{\rm b}=(3/4)~\rho_{\rm ISM}~v_{\rm b}^{2}$.

\noindent To determine the radius of the bow shock, $r_{\rm b}$, we assume a self-similar relationship between its length and width given by $R\equiv l_{\rm b}/r_{\rm b}$, which can change from source to source in the range $R\in [2-3]$ (Leahy et al. \cite{Leahy89}, Kaiser et al. \cite{Kaiser04}, Perucho \& Marti \cite{Perucho07}). Here we use a fiducial value of $R=3$ (similar to the value from our simulations; see Appendix~\ref{appendix}). We assume here a plane shock and do not enter in details concerning the precise geometry of the bow shock. Nevertheless, we need the accelerator size in order to compute the maximum energies that particles can reach, as well as an estimation of the emitter size to provide the surface brightness of the source. In the case of the bow shock, both the accelerator and the emitter size are taken to be $\sim r_{b}$.  The mass density and the pressure, which has been assumed equal to that of the cocoon, have been taken homogeneous in the whole shell. Since the bow shock velocity evolves with time, the mass density and the pressure change as well. Therefore, the magnetic energy density varies also with time. Finally, the radiation field energy density depends on $l_{\rm b}$, $u_{\star}= L_{\star}/4~\pi~l_{\rm b}^{2}~c$ (where $L_{\star}$ is the companion luminosity).

\subsubsection{\it{Cocoon}}

\noindent The cocoon is filled with jet material that crosses the reverse shock. A contact discontinuity separates the cocoon and the shell region materials. We do not consider the mixing of the gas of both zones that could occur due to Kevin Helmholtz and/or Rayleigh Taylor instabilities. 
The cocoon pressure is taken to be the same as that in the other side of the contact discontinuity at any time, $P_{c}=P{b}$. Given the strong compression of the ISM gas in shell zone, the length and the width of the cocoon are taken to be also $\sim l_{\rm b}$ and $\sim r_{\rm b}$, respectively. For the accelerator size, the reverse shock, we adopt the constant jet width after the reconfinement point (see below).
The pressure and mass density of the material can be taken as homogeneous for the whole cocoon to first order approximation. The homogeneity assumption comes from the fact that the cocoon material is expected to be subsonic. Both pressure and mass density will depend on the time evolution of the reverse shock velocity, like the magnetic and radiation fields, in the same way as in the case of the shell.

\subsubsection{\it{Reconfinement region}}

\noindent The conical jet initially emerging from the central engine has an opening angle $\Psi \simeq r/z \sim 0.1$~rad, where $z$ is the distance to the injection point and $r$ is the jet radius. Given the pressure exerted by the surrounding cocoon, the jet radius becomes approximately constant at $z_{\rm conf}$, the distance where the lateral jet pressure becomes similar to the cocoon one. Following Kaiser \& Alexander (1997), we obtain the following radius for the reconfined jet:

\begin{equation}
r_{\rm conf}\sim \Psi \sqrt{\frac{2Q_{\rm jet}v_{\rm jet}}{(\hat{\gamma}+1)(\Gamma_{\rm jet}-1)~\pi c^{2} P_{\rm c}}}\,,
\end{equation}
where $\hat{\gamma}=5/3$ is the adiabatic index of the cocoon material and $v_{\rm jet}$ is the jet velocity. For a recent semi-analytical treatment of reconfinement shocks, see (\cite{Sikora2008}). After injection downstream, the shocked material still moves almost at the jet velocity. The normal component of the velocity of the reconfinement shock, i.e., the one that suffers a discontinuity, is much smaller than $v_{\rm jet}$. The jet keeps most of its thrust and remains supersonic until it is braked at the reverse shock. From the reconfinement point the jet roughly keeps a constant radius. We neglect further recollimation shocks that may occur (see Appendix~\ref{appendix}). 

\noindent The size of the accelerator is taken as the width of the jet at the recollimation shock, $r_{\rm conf}$. Once accelerated, relativistic particles are approximately advected at the jet velocity, and propagate up to the reverse shock. The length of the emitter, assumed to be one-dimensional, is therefore taken to be $\sim l_{\rm b}$ since $z_{\rm conf}\sim r_{\rm conf}/\Psi \ll l_{\rm b}$. The magnetic field in the reconfinement region is derived as in the other interaction zones. The radiation density, however, gradually decays from $z_{\rm conf}$ to $l_{\rm b}$ as the inverse square of the distance to the companion star. The density and the pressure are considered to remain roughly constant after the reconfinement point since the jet radius is constant beyond $z_{\rm conf}$.

\begin{table}[!t]

\caption{List of the parameters which remain with a constant value in the analytic model}

\label{table1}      

\centering                          

\begin{tabular}{l c c}        

\hline\hline                 


{} & {} & {}    \\

\it{Parameter} & \it{Symbol} & \it{Value} \\    

\hline

\\                        

Jet Lorentz factor & $\Gamma_{\rm jet}$ & $1.25$\\ [4pt]

Jet half opening angle & $ \Psi $ & 0.1~rad  \\ [4pt]



Luminosity companion star  & $L_{\star}$ & $10^{39}$ erg~s~$^{-1}$ \\ [4pt]


Self-Similar parameter &  $ R $ & $3$  \\ [4pt]

Magnetic equipartition fraction  & $ \eta $ & $0.1$ \\ [4pt]

Non-thermal luminosity fraction & $ \chi $ & $ 0.01 $ \\ [4pt]

\hline                                   

\end{tabular}

\end{table}



\subsection{The non-thermal particle populations}

\noindent The theory of diffusive shock acceleration in the llinear regime for non-relativistic velocities predicts a power-law index of $p\sim 2$ for the injected particle spectrum in strong shocks (e.g. Drury \cite{Drury83}). We take this value for the non-thermal particle spectra injected at the reconfinement, reverse and bow shock fronts, i.e. $N(E)=K\,E^{-2}$. As noted above, the normalization constant $K$ is 
taken such that $\sim 1$\% of the kinetic power flowing in the jets is converted into non-thermal energy in the postshock region right after each shock. We remark that the radiation luminosities scale linearly with this non-thermal fraction.

\noindent The maximum energies of the relativistic particles, $E_{\rm max}$, are calculated at any source age equating the energy gain to the different cooling processes. In case radiative cooling is not effective, the maximum energy is constrained by the Hillas limit (Hillas \cite{Hillas1984}), i.e. the particle gyroradius, $r_{\rm g}$, equals the accelerator size. We have adopted Bohm diffusion ($D= r_{\rm g}c/3$) and the magnetic field in each shocked region, to calculate the rate at which particles gain energy (e.g. Protheroe \cite{Protheroe1999}):

\begin{equation}
\dot{E}_{\rm accel}(t)\approx (3/20)~[v_{\rm s}(t)/c]^2 e~B(t)~c\,.
\end{equation}

\noindent Focusing hereafter on electrons, the particle energy distribution is computed taking into account radiative (synchrotron, relativistic Bremsstrahlung and IC emission; see, e.g., Blumenthal \& Gould \cite{blum70}) and adiabatic losses along the lifetime of the source. As pointed out in Blundell et al. (\cite{Blundell00}) in their study on extragalactic double radio sources, the spectral aging of the non-thermal populations depends on the evolution of the physical conditions in each region, in particular the expansion velocities and magnetic fields for each interaction zone. At each time, the conditions for the radiative and adiabatic losses change. The non-thermal particle distribution at a given source age $t_{\rm MQ}$ has to be therefore computed summing up the differently evolved contributions of the particles injected previously at the shock fronts all along the source lifetime until reaching $t_{\rm MQ}$.

\noindent For the synchrotron losses, we use the magnetic field considered above for each interaction region. Relativistic Bremsstrahlung is calculated accounting for the densities in the downstream regions. To compute IC losses, we consider the dominant radiation field from the companion, an OB star with luminosity $L_{\star}=10^{39}$~erg~s$^{-1}$. Adiabatic losses, $\dot{E}~(t)~=~[v(t)~/~r~(t)]\,E$, are computed from the size of the emitters $r(t)$ and the expansion velocity $v(t)$. At the downstream region, after the reconfinement shock, there is no expansion since the jet radius keeps roughly constant. In the case of the shell region, the expansion velocity is $v_{\rm b}$, and for the cocoon, it is $\sim 3/4~v_{\rm b}$ (Landau \& Lifshitz \cite{Landau87}). 

\noindent As noted above, the electron population properties, the adiabatic coefficient, and the magnetic and radiation fields are  taken homogeneous as a first order approximation in the bow shock and cocoon regions. A study of the detailed spatial structure of the magnetic field, pressure and mass density at each interaction region is beyond the scope of this work. For the electrons injected at the reconfinement shock, we compute their evolution assuming a one-dimensional emitter, the jet, with the stellar radiation density decaying as $\propto 1/z^2$. 

\noindent The list of the parameters which remain constant in our analytic model are pesented in Table~1.

\noindent Given $E_{\rm max}$ and the evolved electron distribution in each emitting region, and accounting for the mentioned radiation mechanisms, we have computed the SEDs for each zone. This is presented in the next section.

\section{Model Results}\label{res}

\begin{figure}[!tp]

   \centering

   \includegraphics[width=7cm]{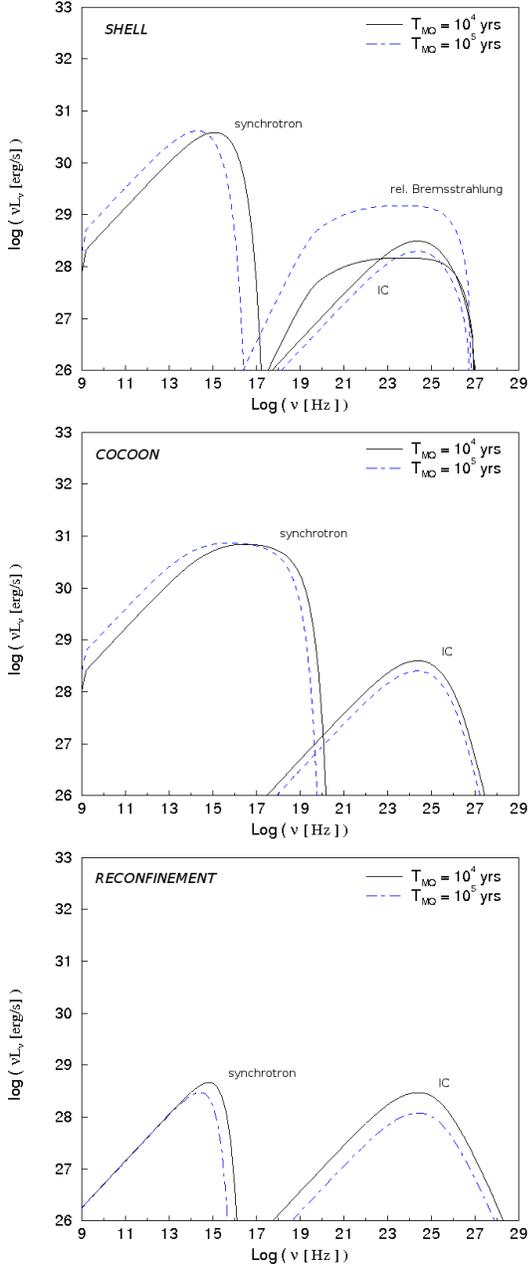} 

   \caption{ Obtained SEDs for the shell (top), cocoon (middle) and jet reconfinement (bottom) regions taking the values for the jet power $Q_{\rm jet}=10^{36}$~erg~s~$^{-1}$ and an external gas particle density $n_{\rm ISM}=0.1$~cm$^{-3}$. Two different values for the source age are represented, $t_{\rm MQ}=$ $10^{4}$~yr (solid line) and $10^{5}$~yr (dashed line). See Table~2 for details on the parameter values for each region.}

         \label{fig2}
   \end{figure}

\begin{figure}[!tp]

   \centering

   \includegraphics[width=7cm]{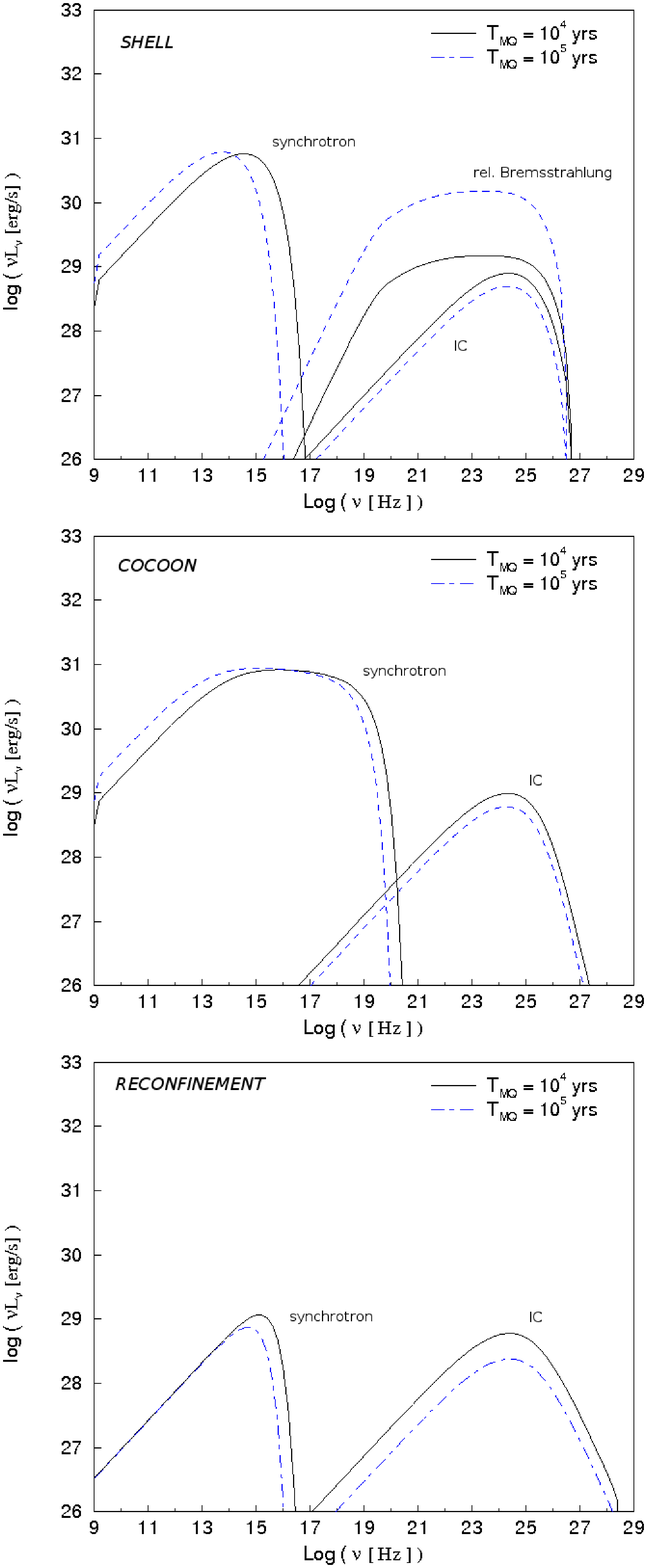}

   \caption{Same as in Fig.~2  but taking $n_{\rm ISM}=1$ cm$^{-3}$. See the physical parameter values in Table~2}

         \label{fig3}

   \end{figure}

\begin{figure}[!tp]

   \centering

   \includegraphics[width=7cm]{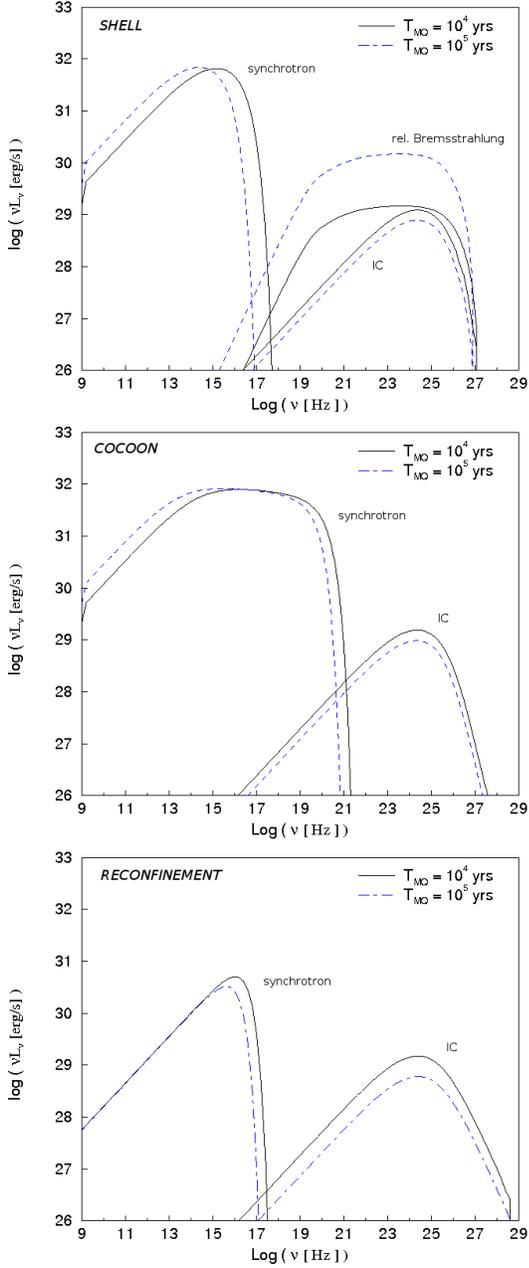}

   \caption{Same as in Fig.~2  but taking  $Q_{\rm jet}= 10^{37}$ erg~s~$^{-1}$. See the physical parameter values in Table~2}

         \label{fig4}

   \end{figure}

\begin{figure}[!tp]

   \centering

   \includegraphics[width=7cm]{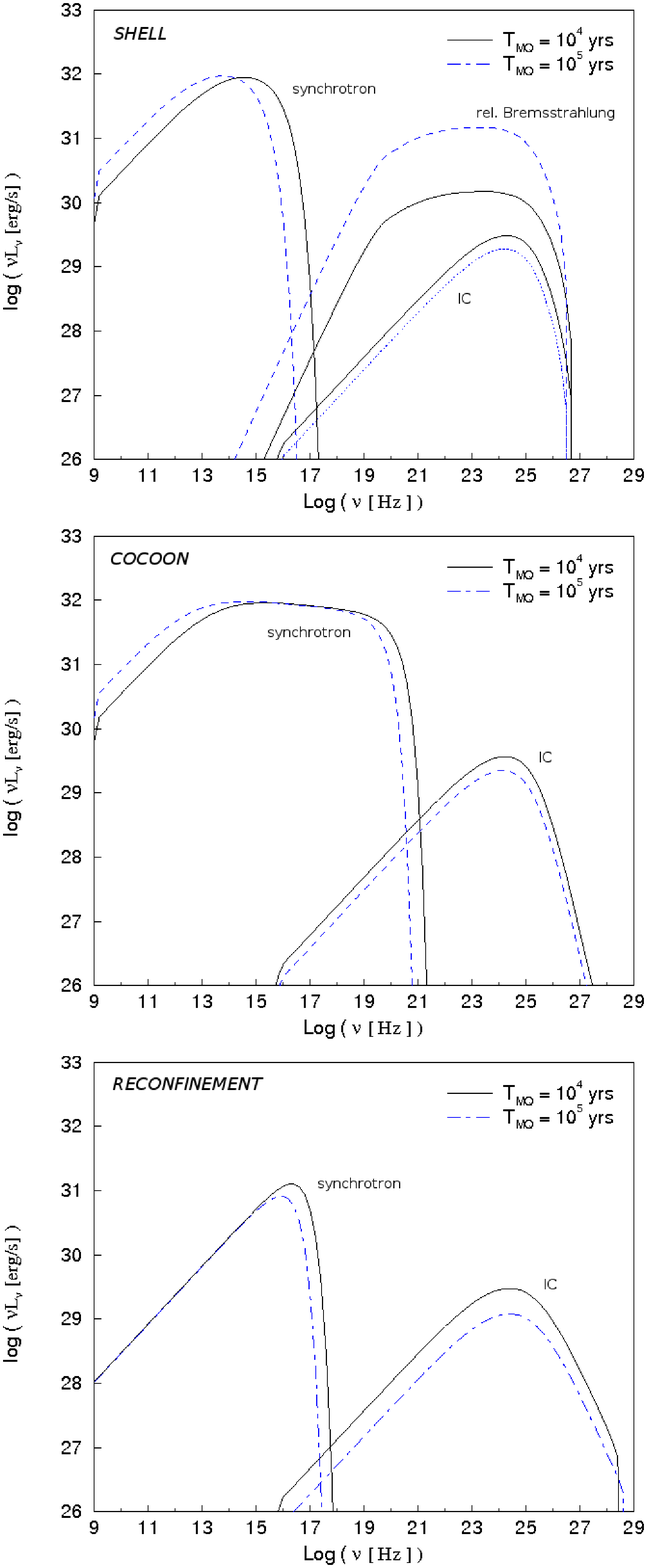}

   \caption{Same as in Fig.~2  but taking  $Q_{\rm jet}= 10^{37}$ erg~s~$^{-1}$ and $n_{\rm ISM}=1$~cm$^{-3}$. See physical parameter values in Table~2}

         \label{fig5}

   \end{figure}

\noindent We have studied how the computed SEDs are affected by varying the source age, $t_{\rm MQ}$, the jet kinetic power, $Q_{\rm jet}$, and the ISM density, $n_{\rm ISM}$. In Figs.~\ref{fig2} to \ref{fig5} we show the SEDs for the shell (top), cocoon (middle) and jet reconfinement (bottom) regions. The contribution of only one jet interacting with the ISM is accounted. The whole set of parameter values, required for the different cases to obtain the particle evolution and subsequent emission, are presented in Table~2.

\noindent Synchrotron emission from the three interaction zones is the channel through which the highest radiation output is obtained, with bolometric luminosities that can be as high as $\sim~10^{33}$~erg~s$^{-1}$ for powerful sources (see Figs.~4 and 5). At high and very high energies, IC emission is the dominant process in the cocoon and reconfinement regions, reaching luminosities up to $\sim 10^{30}$~erg~s$^{-1}$, while for the shell zone relativistic Bremsstrahlung dominates at this energy range, with luminosities up to $\sim~10^{32}$~erg~s$^{-1}$. Notable differences are found in the reported SEDs when varying the source age $t_{\rm MQ}$ from $10^{4}$ to $10^{5}$ yr. For older sources, the interaction zones are located at larger distances from the companion star, its radiation energy density $u_{\star}$ decreases and the IC contribution gets slightly lower, although particle aging leads to higher emission at lower energies in the shell and cocoon regions. Relativistic Bremsstrahlung in the shell zone is strongly affected by the source age, due to a particle accumulation effect, giving luminosities a factor of $\sim 10$ larger for old sources. Regarding the ambient medium density, higher values of $n_{\rm ISM}$ make the jet to be braked at shorter distances from the central engine. The interaction regions are then under a higher radiation energy density from the companion star, and the IC emission is accordingly enhanced. The relativistic Bremsstrahlung emission in the shell zone is also higher for denser mediums, since the luminosity is proportional to the target ion field density, $n_{\rm t}=4\,n_{\rm ISM}$. Finally, we note that in our model all the non-thermal luminosities are proportional to $Q_{\rm jet}$.


\begin{table*}[!tp]

\caption{Parameter values adopted to compute the SEDs for the three emitting zones. The target density, $n_{\rm t}$, is only shown when non-negligible. We show the parameter values at $t_{\rm MQ}$, but we remark that they smoothly vary with time. The computation of the non-thermal particle distribution have taken this into account. See Section~2 for details. }


\label{table2}                     

\centering                          

\begin{tabular}{| l | c | c | c | c | c | c | c | c |} 

\hline 

\hline



 & \multicolumn{2}{c|}{} & \multicolumn{2}{c|}{} & \multicolumn{2}{c|}{} & \multicolumn{ 2}{c|}{} \\ 

\it{Parameter} & \multicolumn{2}{c|}{Fig. 2} & \multicolumn{2}{c|}{Fig. 3} & \multicolumn{2}{c|}{Fig. 4} & \multicolumn{ 2}{c|}{Fig. 5} \\ [1pt] \hline 

\hline

 & \multicolumn{4}{c|}{} & \multicolumn{4}{c|}{}  \\ 

Jet kinetic power $Q_{\rm jet}$ (erg~s~$^{-1}$ ) & \multicolumn{4}{c|}{$10^{36}$} & \multicolumn{4}{c|}{$10^{37}$} \\ [5pt]

\hline

 & \multicolumn{2}{c|}{} & \multicolumn{2}{c|}{} & \multicolumn{2}{c|}{} & \multicolumn{ 2}{c|}{} \\ 

ISM density $n_{\rm ISM}$ (cm$^{-3}$)      & \multicolumn{2}{c|}{$0.1$} & \multicolumn{2}{c|}{$1$} & \multicolumn{2}{c|}{$0.1$} & \multicolumn{2}{c|}{$1$} \\ [5pt]

\hline

 & {} & {} & {} & {} & {} & {} & {} & {} \\ 

Source age $t_{\rm MQ}$ (yr)    & $10^{4}$ & $10^{5}$ & $10^{4}$ & $10^{5}$ & $10^{4}$ & $10^{5}$ & $10^{4}$ & $10^{5}$ \\ [5pt]

  \hline

  & \multicolumn{7} {}{} & {} \\

 {\it SHELL}  & \multicolumn{7} {}{} & {} \\

  & \multicolumn{7} {}{} & {}\\

  \hline

 & {} & {} & {} & {} & {} & {} & {} & {} \\

Magnetic field $B$ (G)                     & 2.9 $\times 10^{-5}$ & 1.1 $\times 10^{-5}$ & 5.8 $\times 10^{-5}$ & 2.3 $\times 10^{-5}$ & 4.6 $\times 10^{-5}$ & 1.8 $\times 10^{-5}$ & 9.2 $\times 10^{-5}$ & 3.6 $\times 10^{-5}$  \\ [4pt]

Shock velocity $v_{\rm b}$ (cm s$^{-1}$)       & 4.3 $\times$ 10$^{7}$ & 1.7 $\times$ 10$^{7}$  & 2.7 $\times$ 10$^{7}$  & 1.0 $\times$ 10$^{7}$  & 6.9 $\times$ 10$^{7}$  & 2.7 $\times$ 10$^{7}$  & 4.3 $\times$ 10$^{7}$  & 1.7 $\times$ 10$^{7}$   \\ [4pt]

Emitter size $r$ (cm)             & 2.3 $\times 10^{19}$ & 9.1 $\times 10^{19}$ & 1.4 $\times 10^{19}$ & 5.7 $\times 10^{19}$ & 3.62 $\times 10^{19}$ & 1.44 $\times 10^{20}$ & 2.3 $\times 10^{19}$ & 9.1 $\times 10^{19}$  \\ [4pt]

Rad. energy dens. $u_{\star}$ (erg cm$^{-3}$) & 5.0 $\times 10^{-12}$ & 3.2 $\times 10^{-11}$ & 1.2 $\times 10^{-13}$ & 8.0 $\times 10^{-13}$ & 2.0 $\times 10^{-12}$ & 1.3 $\times 10^{-12}$ & 5.0 $\times 10^{-13}$ & 3.2 $\times 10^{-13}$  \\ [4pt]

Maximum energy $E_{\rm max}$  (TeV)              & 8.1 & 5.1 & 3.6 & 2.3 & 10.2 & 6.4 & 4.5 & 2.8  \\ [4pt]

Target density  $n_{\rm t}$ (cm$^{-3}$)    & 0.4  & 0.4  & 4.0  & 4.0  & 0.4  & 0.4  & 4.0  & 4.0  \\ [4pt]

  \hline

  & \multicolumn{7} {}{} & {} \\

 {\it COCOON}    & \multicolumn{7} {}{} & {} \\

  & \multicolumn{7} {}{} & {} \\

  \hline

 & {} & {} & {} & {} & {} & {} & {} & {} \\

Magnetic field $B$ (G)                         & 3.8 $\times 10^{-5}$ & 1.5 $\times 10^{-5}$ & 7.5 $\times 10^{-5}$ & 3.0 $\times 10^{-5}$ & 5.9 $\times 10^{-5}$ &  2.4 $ \times 10^{-5}$ & 1.2  $\times 10^{-4}$  & 4.7 $\times 10^{-5}$ \\ [2pt]

Shock velocity $v_{\rm s}$ (cm s$^{-1}$)   & 1.8 $\times 10^{10}$ & 1.8 $\times 10^{10}$ & 1.8 $\times 10^{10}$ & 1.8 $\times 10^{10}$ & 1.8 $\times 10^{10}$ & 1.8 $\times 10^{10}$ & 1.8 $\times 10^{10}$ & 1.8 $\times 10^{10}$ \\ [4pt]

Emitter size $r$ (cm)               & 6.5 $\times 10^{18}$ & 2.5 $\times 10^{18}$ & 4.0 $\times 10^{19}$ & 1.6 $\times 10^{19}$ & 1.0 $\times 10^{19}$ & 4.0 $\times 10^{18}$ & 6.5 $\times 10^{19}$ & 2.5 $\times 10^{19}$ \\ [4pt]

Rad. energy dens. $u_{\star}$ (erg cm$^{-3}$)   & 7.0 $\times 10^{-12}$ & 1.7 $\times 10^{-11}$ & 4.4 $\times 10^{-13}$ & 1.1 $\times 10^{-12}$ & 2.8 $\times 10^{-12}$ & 1.7 $\times 10^{-12}$ & 7.0 $\times 10^{-13}$ & 4.4 $\times 10^{-13}$ \\ [4pt]

Maximum energy $E_{\rm max}$ (TeV)                & 275.5 & 275.5 & 275.5 & 275.5 & 871.2 & 871.2 & 871.23 & 871.2 \\ [4pt]

  \hline

  & \multicolumn{7} {}{} & {} \\

 {\it RECONFINEMENT}  & \multicolumn{7} {}{} & {} \\

  & \multicolumn{7} {}{} & {} \\

  \hline

 & {} & {} & {} & {} & {} & {} & {} & {} \\

Magnetic field $B$ (G)                         & 2.6 $\times 10^{-5}$ & 1.0 $\times 10^{-5}$ & 5.2 $\times 10^{-5}$ & 2.1 $\times 10^{-5}$ & 4.2 $\times 10^{-5}$ & 1.6 $\times 10^{-5}$ & 8.3 $\times 10^{-5}$ & 3.3 $\times 10^{-5}$ \\ [4pt]

Shock velocity $v_{\rm conf}$ (cm s$^{-1}$)          & 1.8 $\times 10^{9}$ & 1.8 $\times 10^{9}$ & 1.8 $\times 10^{9}$ & 1.8 $\times 10^{9}$ & 1.8 $\times 10^{9}$ & 1.8 $\times 10^{9}$ & 1.8 $\times 10^{9}$ & 1.8 $\times 10^{9}$ \\ [4pt]

Emitter size $r$ (cm)               & 1.9 $\times 10^{19}$ & 7.6 $\times 10^{19}$ & 1.2 $\times 10^{19}$ & 4.8 $\times 10^{19}$ & 3.0 $\times 10^{19}$ & 1.2 $\times 10^{20}$ & 1.9 $\times 10^{19}$ & 7.6 $\times 10^{19}$ \\ [4pt]

Rad. energy dens. $u_{\star}$ (erg cm$^{-3}$)  & 2.5 $\times 10^{-8}$ & 4.0 $\times 10^{-9}$ & 1.0 $\times 10^{-7}$ & 1.6 $\times 10^{-8}$ & 6.4 $\times 10^{-9}$ &  1.0 $\times 10^{-9}$ & 2.5 $\times 10^{-8}$ & 4.0 $\times 10^{-9}$ \\ [4pt]

Maximum energy $E_{\rm max}$ (TeV)                 & 3.2 & 3.2 & 3.2 & 3.2 & 10.1 & 10.1 & 10.1 & 10.1 \\ [4pt]

\hline                                   

\end{tabular}

\end{table*}



\subsection{Emission from the shell region}

\noindent Under the acceleration conditions in this region, the maximum energies $E_{\rm max}$ of electrons for the different cases range 2--10 TeV, being always limited by synchrotron losses. Synchrotron emission is also the dominant radiative channel, peaking at  higher frequencies for higher $E_{\rm max}$. The electron energy distribution has an energy break, $E_{\rm b}$, at which the radiative cooling timescale and the age of the source are the same. $E_{\rm b}$ changes its value depending on the physical conditions: older sources render smaller $E_{\rm b}$, since particles have cooled via synchrotron for a longer time; higher densities yield lower $E_{\rm b}$, due to smaller $l_{\rm b}$ and therefore larger $B$ and synchrotron cooling. Larger jet powers enhances $B$ as well, reducing again $E_{\rm b}$ (this discussion about $E_{\rm b}$ is valid also for the other interaction regions). The energy break is manifested in the synchrotron spectrum at frequencies in the range $\sim 10^{14}-10^{16}$~Hz. Given the long synchrotron cooling timescales $\tau_{\rm sync}\sim 10^{14} B_{-4}^{-2} E_{\rm b~GeV}^{-1}$~s ($B_{-4}=B/10^{-4}$~G) compared with the age of the source $\sim 10^{12}$~s, only the highest energy band of the particle spectrum has reached the steady state. The synchrotron luminosities reach values up to $\sim 10^{33}$~erg~s$^{-1}$ (see Figs.~\ref{fig4} and \ref{fig5}). The Relativistic Bremsstrahlung has a strong dependence on the density, follows the energy distribution of electrons, and can reach luminosities $\sim 10^{32}$~erg~s$^{-1}$ in the 1~MeV--10~GeV range (see Fig.~\ref{fig5}). The IC luminosities are well below the synchrotron ones in all the studied cases  .

\subsection{Emission from the cocoon}

\noindent Electrons here can reach energies of about  $E_{\rm max} \sim$~100~TeV, being limited by escape losses. Synchrotron emission, again the dominant radiative channel, extends up to hard X-rays in the cocoon, with luminosities reaching up to $\sim 10^{33}$~erg~s$^{-1}$ (see Fig.~\ref{fig5}) depending on the different parameter values. As in the shell region, the IC luminosities are well below the synchrotron ones. Otherwise, due to the low densities in the cocoon, relativistic Bremsstrahlung is negligible in this zone.

\subsection{Emission from the jet reconfinement region}

\noindent The size of the accelerator (Hillas \cite{Hillas1984}) put a limit of $\sim 10$~TeV on the maximum energies that electrons can acquire at the reconfinement shock. An increase in both the synchrotron and the IC emissivities is found for denser mediums, in which case $P_{\rm c}$ is larger yielding smaller $z_{\rm conf}$, which determines the magnetic and radiation field energy densities. The synchrotron luminosity is similar to the IC one for weak jets, whereas it is well above it for powerful jets, where luminosities $\sim 10^{32}$~erg~s$^{-1}$ are obtained (see Figs.~4 and 5). This is due to the stronger dependence of the synchrotron emission with the jet power. The stellar radiation field enhances the IC contribution when the set of parameters make $z_{\rm conf}$ shorter. As in the cocoon region, the relativistic Bremsstrahlung contribution is negligible. 

\section{Hydrodynamical simulations \textit{vs} analytical model}

\noindent We performed hydrodynamical simulations to further study the interaction of MQ jets with their surroundings (see Appendix for details). The values of $Q_{\rm jet}$, $t_{\rm MQ}$ and $n_{\rm ISM}$ adopted in the hydrodynamical simulations are similar to those assumed in the analytical model, so it is worthwhile checking the main physical quantities used to compute the non-thermal emission. The pressure evolution with simulation time have been computed for the shell and cocoon regions. We show also a fit of their evolution performed to check their values up to $t_{\rm MQ} = 10^{5}$~yr. From an initial value of $\sim~2 \times 10^{-9}$ erg cm$^{-3}$, the cocoon pressure smoothly reaches $\sim~8~\times~10^{-11}$~erg~cm$^{-3}$ at t~$= 7 \times 10^{11}$~s. The shell behaves similarly, as expected, with a final pressure of $\sim~8~\times~10^{-11}$~erg~cm$^{-3}$. These values are in reasonable agreement with those found in the analytical model, ranging between $(2-7)\times 10^{-10}$~erg~cm$^{-3}$. Concerning the mass densities, $\rho_{\rm ISM}$ stabilizes at $\sim 2 \times 10^{-25}$~g~cm$^{-3}$ and $\rho_{\rm co} \sim 4 \times 10^{-29}$~g~cm$^{-3}$ for the shell and the cocoon regions, respectively. These values are again very similar to those found in the analytical treatment. On the other hand, the geometry of the emitting structures is related to the self-similar ratio $R$. In the analytical model, we use $R=3$. We find this value to be in accordance with the results of the numerical simulations, which yield a value between 2.5~and~2.7.~Finally, we note that in our model only a strong shock at the reconfinement point is assumed, and no additional shocks are considered along the jet. Otherwise, the hydrodynamical simulations show the existence of several conical shocks that develop in the jet when its pressure falls to that of the surrounding cocoon (see Fig.~\ref{shocks}). Therefore, the non-thermal emission presented in Figs.~\ref{fig2} to \ref{fig5} for the reconfinement region, with only one strong shock and an acceleration efficiency of $\chi=0.01$, should be taken as a rough approximation of the real situation.

\section{Discussion}

\noindent We have explored whether non-thermal emission can be expected from the bow shock, the cocoon and the reconfinement regions in the interaction of MQ jets with their environment. In the extragalactic framework, non-thermal radiation is supposed to come only from the cocoon as extended radio emission, as well as locally in the hot spots at the jet tips. Although the shell of shocked ambient material plays an important role in the analytical models describing the growing of FR~II galaxies (Blandford et al. \cite{blandford1974}, Scheuer \cite{Scheuer1974}), no radio emission is usually assumed to be produced there (see, however, Rudnick et al. \cite{Rudnick1988}). In our model, the bow shock velocity is still large enough to accelerate particles, so it is worth to compute the expected non-thermal emission also from this region. Our results show a contribution from this zone that is comparable to that of the cocoon and higher than that coming from the reconfinement region. The geometry of the interaction structures, however, makes it difficult to disentangle the emission from the cocoon and the bow shock region since they could appear co-spatial in the plane of the sky. A clear relativistic Bremsstrahlung would favour a shell origin of the emission.

\noindent The highest radiation output within the studied set of parameters corresponds to the case: $Q_{\rm jet}=10^{37}$~erg~s$^{-1}$, $t_{\rm MQ}=10^{5}$~yr and $n_{\rm ISM}=1$~cm$^{-3}$. In case the emitting source were located at 3 kpc, this would imply a flux density of $\sim 150$~mJy at 5~GHz. The emitting size would be of a few arcminutes, since the electron cooling timescale is longer than the source lifetime and they can fill the whole cocoon/shell structures. Considering this angular extension and taking a radio telescope beam size of $10''$, radio emission at a level of $\sim 1$~mJy~beam$^{-1}$ could be expected. At the X-ray band, we find a bolometric flux in the range 1--10~keV of $F_{\rm 1-10 keV}$ $\sim 2 \times 10^{-13}$~erg~s$^{-1}$~cm$^{-2}$. The electrons emitting at X-rays by synchrotron have very short time-scales, and the emitter size cannot be significantly larger than the accelerator itself. Although the X-rays produced in the shell through relativistic Bremsstrahlung are expected to be quite diluted, the X-rays from the cocoon would come from a relatively small region close to the reverse shock, and could be detectable by {\it XMM-Newton} and {\it Chandra} at scales of few arcseconds. In the gamma-ray domain, the flux between 100 MeV and 100 GeV is $F_{\rm 100~MeV<E<100~GeV}\sim 10^{-14}$~erg~s$^{-1}$~cm$^{-2}$, while the integrated flux above 100~GeV is $F_{\rm E>100GeV}\sim 10^{-15}$~erg~s$^{-1}$~cm$^{-2}$. These values are too low to be detectable by current Cherenkov telescopes. Taking into account the rough linearity between $Q_{\rm jet}$, $n_{\rm ISM}$, $t_{\rm MQ}$, $\chi$ and $d^{-2}$ with the gamma-ray fluxes obtained, sources with higher values of these quantities than the ones used here may render the MQ jet termination regions detectable by present Cherenkov telescopes. For the weakest jets, i.e, lowest ISM densities and youngest sources adopted in our model ($Q_{\rm jet}=10^{36}$~erg~s$^{-1}$, $t_{\rm MQ}=10^{4}$~yr and $n_{\rm ISM}=0.1$~cm$^{-3}$), the fluxes are strongly suppressed. In the radio band, the specific flux is $F_{\rm 5~GHz}\sim 0.1$~mJy~beam$^{-1}$; the integrated flux at X-rays in this case is $F_{\rm 1-10~keV}\sim 10^{-14}$~erg~s$^{-1}$~cm$^{-2}$, and at gamma-rays $F_{\rm 100MeV<E<100GeV}\sim 2\times 10^{-17}$~erg~s$^{-1}$~cm$^{-2}$ and $F_{\rm E>100GeV}\sim$ $2\times 10^{-16}$~erg~s$^{-1}$~cm$^{-2}$.

\noindent We remark that the fluxes showed above strongly depend on the non-thermal luminosity fraction $\chi$. In the present model we use a quite conservative value of $\chi=0.01$. In the case of a source able to accelerate particles at a higher efficiency at the interaction shock fronts, then the expected non-thermal fluxes would be enhanced by a factor ($\chi / 0.01$).

\noindent The non-thermal emission from the interaction regions in MQs presents characteristic features that can be distinguished from emission coming from the central binary system. First of all, the interaction structures are localized at distances up to $\sim~10^{20}$~cm. The lifetime of electrons radiating at high and very high energies is relatively short thus the emission region should be localized near the accelerator (but far away from the central system). Regarding radio emission, synchrotron cooling times are expected to be larger than the source lifetime, $t_{\rm MQ}=10^{4}$ - $10^{5}$~yr. The cocoon and bow shock emitting regions would be expected to form a kind of radio nebula around the central system and the flux densities at the level showed above would then come from a quite extended source.

\noindent The detection of non-thermal emission from the interaction zones would be a proof that efficient acceleration of particles up to relativistic energies is taking place far away from the central binary system. The averaged kinetic power carried away in the jets could be better constrained, eventually showing that it can be much higher than that inferred directly from observations of the inner jet emission alone (Gallo et al. \cite{gallo05}; Heinz \cite{heinz06})

\noindent Despite we focus on the non-thermal emission from the MQ jet termination regions, thermal Bremsstrahlung emission should be expected from the shell. Although the shocks considered here are still adiabatic, a non-negligible fraction of the jet kinetic luminosity of up to a few \% may be radiated via thermal Bremsstrahlung. For bow-shock velocities of few times $10^7$~cm~s$^{-1}$, the thermal emission would peak at UV-soft X-rays, energy band that is strongly affected by absorption in the ISM. Observations of the thermal radiation in radio and optical from the interaction structures have been used to extract information of the shell physical conditions (e.g. Cygnus X-1, Gallo et al. \cite{gallo05}; Russell et al. \cite{Russell07}).

\noindent The role of accelerated protons in the shocks deserves a few words, since it may be relevant in some specific situations. Given the conditions in the strong shocks we are considering, relativistic protons may reach energies of about 100 TeV; for shell densities $\sim$~1~cm$^{-3}$, the accelerated protons have lifetimes of about 10$^{15}$~s. To reach the gamma-ray fluxes detectable for the present generation of satellite borne and ground based gamma-ray instruments above $\sim 100$~GeV, $\sim 10^{-13}$~erg~cm$^{-2}$~s$^{-1}$, the energy in non-thermal protons stored in the shell should be as high as $\sim 10^{48}$~erg at few kpc distances. For a source age of $t_{\rm MQ}=10^{5}$~yr, the required injected power in relativistic protons should be about $\sim 3\times 10^{35}$~erg~s$^{-1}$, thus implying that moderate levels of hadronic emission from MQ jet termination regions may be eventually detected from powerful jets, i.e. $Q_{\rm jet} \gtrsim 10^{37}$~erg~s$^{-1}$.

\noindent The reason why some MQs show non-thermal emission from the jet/ISM interaction regions, whereas in other cases such emission remains undetected, is still unclear. In the context of our model, we can study the effects of varying the set of parameters defining the source and their environment, and predict some cases in which the interaction structures may or may not be detectable. First of all, the energy input injected to the medium should be high enough, and there exist strong differences in the jet kinetic power from source to source. In addition, it could be also the case that the density of the surrounding medium is so low that the shell and the cocoon get very large and their radiation too diluted to be detectable (see, for instance, Heinz \cite{Heinz02}). Moreover, MQ jets could get disrupted at some source age, as it is found in FR~I galaxies. If this happened within times much shorter than the MQ lifetime, the probability to detect a cocoon/shell structure in the MQ surroundings would be smaller (although the detection of some other kind of structures is not discarded). On the other hand, some sources may be too far, or the non-thermal fraction too small, to detect significant emission from the interaction regions.

\noindent The evolution of the pressure, mass density, the velocities and the Mach number predicted by the analytical model are in good agreement with those found in the hydrodynamical simulations for the shell and the cocoon regions. Otherwise, these simulations show that several conical shocks may be present within the jet as a consequence of pressure adjustments with the surrounding cocoon, instead of the one strong shock adopted in the analytical treatment. Finally, the length and width of the structures in the model and those found through the numerical simulations are also similar, with a constant ratio $R\sim 3$ in both cases, implying that the physical assumptions used in the analytical treatment are valid to first order.

\noindent The results of this work show that the surroundings of some MQs could be extended non-thermal emitters from radio to gamma-rays. In addition, from a comparison with observations, the magnetic field and the particle acceleration efficiency in the jet/ISM interaction regions can be constrained, giving an insight on the physics of these interaction structures. It is interesting to note that, although the adopted model is rather simple, it already accounts for cases when the sources should remain undetectable and cases in which radiation could be detected.


\begin{acknowledgements}

The authors acknowledge support by the Spanish DGI of MEC under grant AYA2007-6803407171-C03-01, as well as partial support by the European Regional Development Fund (ERDF/FEDER). P.B. was supported by the DGI of MEC (Spain) under fellowship BES-2005-7234. V.B-R. gratefully acknowledges support from the Alexander von Humboldt Foundation. MP acknowledges support from a postdoctoral fellowship of the \emph{Generalitat Valenciana} (\emph{Beca Postdoctoral d'Excel$\cdot$l\`encia}), a Max-Planck-Institut postdoctoral fellowship and by the Spanish MEC and the European Fund for Regional Development through grants AYA2007-67627-C03-01 and AYA2007-67752-C03-02.

\end{acknowledgements}

 \newpage
 \balance

 \newpage

\begin{appendix}
\section{ hydrodynamical simulations}\label{appendix}

The simulation has been performed in order to check the physical values adopted in the analytical model. We used a two-dimensional finite-difference code based on a high-resolution shock-capturing scheme, which solves the equations of relativistic hydrodynamics written in conservation form. This code is an upgrade of the code described in Mart\'{\i} et al. (\cite{mart97}) (see Perucho et al. \cite{pe+05}). The simulation has been carried out in two dual-core processors in the Max-Planck-Institut f\"ur Radioastronomie.

The numerical grid of the simulation is formed by 320 cells in the radial direction and 2400 cells in the axial direction in a uniform region, with physical dimensions of 40$\times\,600$ $r_{\rm j}$. An expanded grid with 320 cells in the transversal direction, brings the boundary from $40\,r_{\rm j}$ to $500\,r_{\rm j}$, whereas an extended grid in the axial direction, consisting of 440 extra cells, spans the grid axially from $600\,r_{\rm j}$ to $900\,r_{\rm j}$. The enlargement of the grid is necessary to ensure that the boundary conditions are sufficiently far from the region of interest, and to avoid numerical reflection of waves in the boundaries affecting our results. The conditions at the boundaries are reflection on the jet axis and in the side where the jet is injected, simulating the presence of the counter-jet cocoon, with the exception of the injection point, where inflow conditions are used. Finally, outflow conditions in the outer axial and radial boundaries are used.

The numerical resolution in the uniform grid is thus of 8 cells/$r_{\rm j}$ in the radial direction and 4 cells/$r_{\rm j}$ in the axial direction. The low resolution used is justified by the fact that we are interested in the macroscopic features of the jet and backflow, but not in mixing and turbulence, allowing much less time consuming simulations. All the physical variables are scaled to the units of the code, which are the jet radius $r$, the rest-mass density of the ambient medium $\rho_{\rm ISM}$, and the speed of light $c$.

The jet is injected in the grid at a distance of $10^{18}$~cm from the compact object, and its initial radius is taken to be $r_0=10^{17}$~cm. The time unit of the code is thus equivalent to $\approx 3\times 10^6$~s, which is derived using the radius of the jet at injection and the speed of light ($r_0/c$). Both the jet and the ambient medium are considered to be formed by a non-relativistic gas with adiabatic exponent $\Gamma=5/3$. The number density in the ambient medium is $n_{\rm ISM}=0.3$~cm$^{-3}$ and the temperature is $T=100$~K. The role of the temperature is not relevant in our context. The velocity of the jet at injection is $0.6\,c$, its number density $n_{\rm j}=1.4\times 10^{-5}\,\rm{cm^{-3}}$, and temperature $T\sim 10^{11}$~K (which corresponds to a sound speed $\sim 0.1\,v_{\rm jet}$). These parameter values result in a jet power $Q_{\rm jet}=3\times 10^{36}$~erg~s$^{-1}$. Figs.~\ref{vmach} and \ref{rhopres} show the velocity, Mach number, mass density and pressure maps obtained with the numerical simulations. The upper (lower) panels in Fig.~\ref{pressions_i_densitats} show the evolution of the pressure and the mass density with simulation time for the shell (cocoon) region. The evolution of the self-similar parameter $R$ reaches a value between 2.5 and 2.7 at the end of the simulations, as can be seen in Fig.~\ref{ssimilar}.

\begin{figure}[b]

   \centering

   \includegraphics[width=8cm]{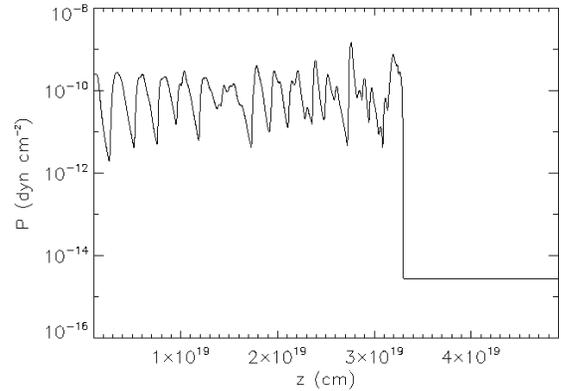}

      \caption{Pressure along the jet axis as a function of distance from its injection point, at $\sim 10^{18}$~cm, as found in the hydrodinamical simulations. Several conical shocks are present, due to the pressure balance with the surrounding cocoon: each time the jet pressure falls below that of the cocoon, a shock is formed, keeping the jet radius roughly constant until it reaches the reverse shock.}

         \label{shocks}

   \end{figure}

\begin{figure}[b]

   \centering

   \includegraphics[width=8cm]{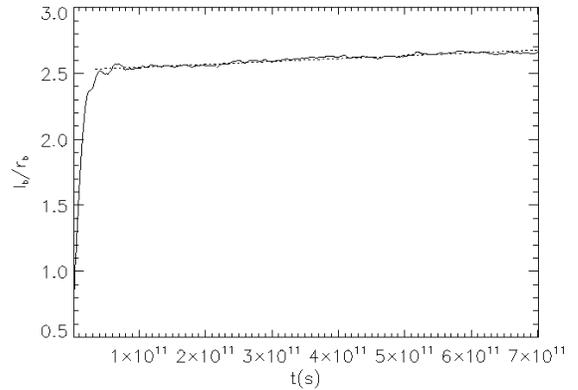}

      \caption{Evolution of the self-similar parameter $R=l_{\rm b}/r_{\rm b}$ as a function of time. After the pronounced initial increase, $R$ remains between 2.5 and 2.7 for most of the simulated time.} 

     \label{ssimilar}

   \end{figure}

At the time the simulation is stopped, after evolving $\approx 2.3\times 10^4$~yr, the bow-shock is moving at a speed $\approx 3\times 10^7\,\rm{cm s^{-1}}$, and has reached a distance $\sim 3.3\times 10^{19}$~cm. Initially, the jet expands, accelerating and cooling, due to its initial overpressure. When the flow becomes underpressured with respect to the ambient medium, a first reconfinement shock is generated close to the injection $2\times 10^18$~cm. The fluid becomes then again overpressured when passing through the shock and this process is repeated several times around pressure equilibrium with the external medium. The subsequent shocks produced by these oscillations around equilibrium are stronger than the first, suggesting some coupling to a pinching Kelvin-Helmholtz instability. The jet is supersonic at injection, with Mach number $M_{\rm j}=6.5$ on axis, before the first reconfinement shock. After this shock, the Mach number oscillates around the initial value and decreases, with slight increases in the expansion regions, until the head of the jet. Here, transonic and subsonic velocities result from the increase in temperature and decrease in velocity, as the flow crosses the reverse shock. The cocoon and the shell material are still in high overpressure with respect to the ambient by the end of the simulation, in a factor $>10^4$. The gas that forms the cocoon is initially fast and slightly supersonic, with velocities up to $9\times 10^9\,\rm{cm s^{-1}}$ and Mach numbers up to 2, close to the head of the jet. Farther downstream, the backflow gets slower and subsonic, with velocities $<6\times 10^8\,\rm{cm s^{-1}}$.

   \begin{figure*}[!b]

   \centering

   \includegraphics[width=15cm]{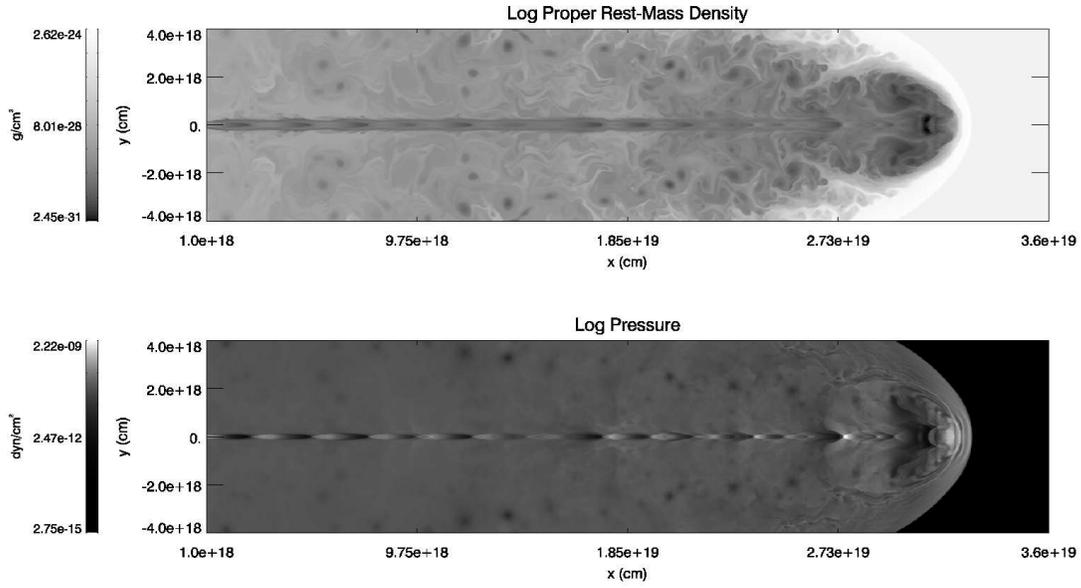}

      \caption{Mass density (top) and pressure (bottom) maps resulting from hydrodynamical simulations. The simulations were performed using $Q_{\rm jet}=3 \times 10^{36}$~erg~s$^{-1}$, $t_{\rm MQ}=3 \times 10^{4}$~yr and $n_{\rm ISM}$=0.3~cm$^{-3}$.}
         \label{rhopres}
   \end{figure*}

  \begin{figure*}[!tp]

   \centering

   \includegraphics[width=15cm]{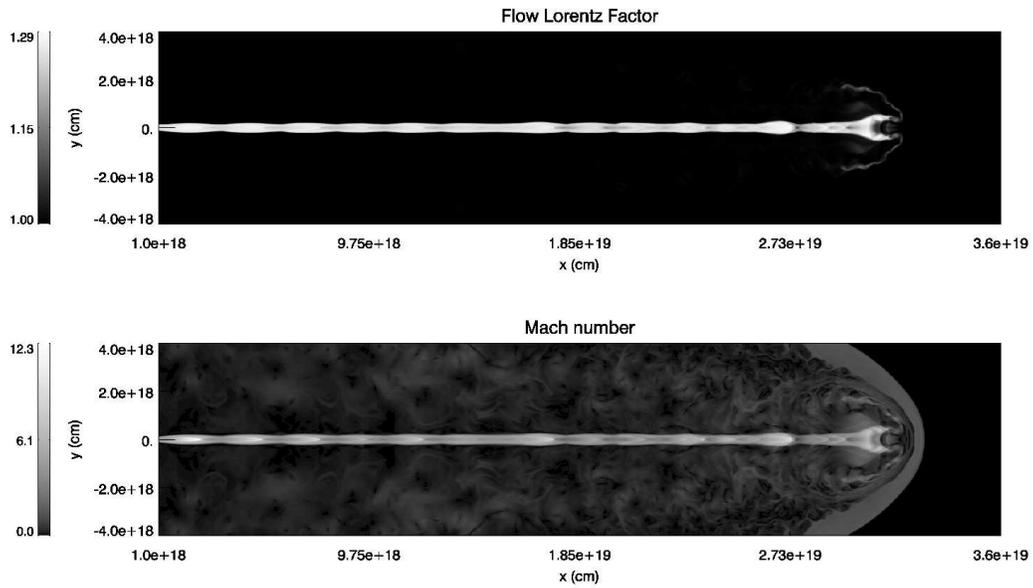}

      \caption{Lorentz factor (top) and Mach number (bottom) maps resulting from hydrodynamical simulations. The adopted parameters are the same as those of Fig.~\ref{rhopres}.}
      \label{vmach}

   \end{figure*}

 \begin{figure*}[!tp]

   \centering

   \includegraphics[width=15cm]{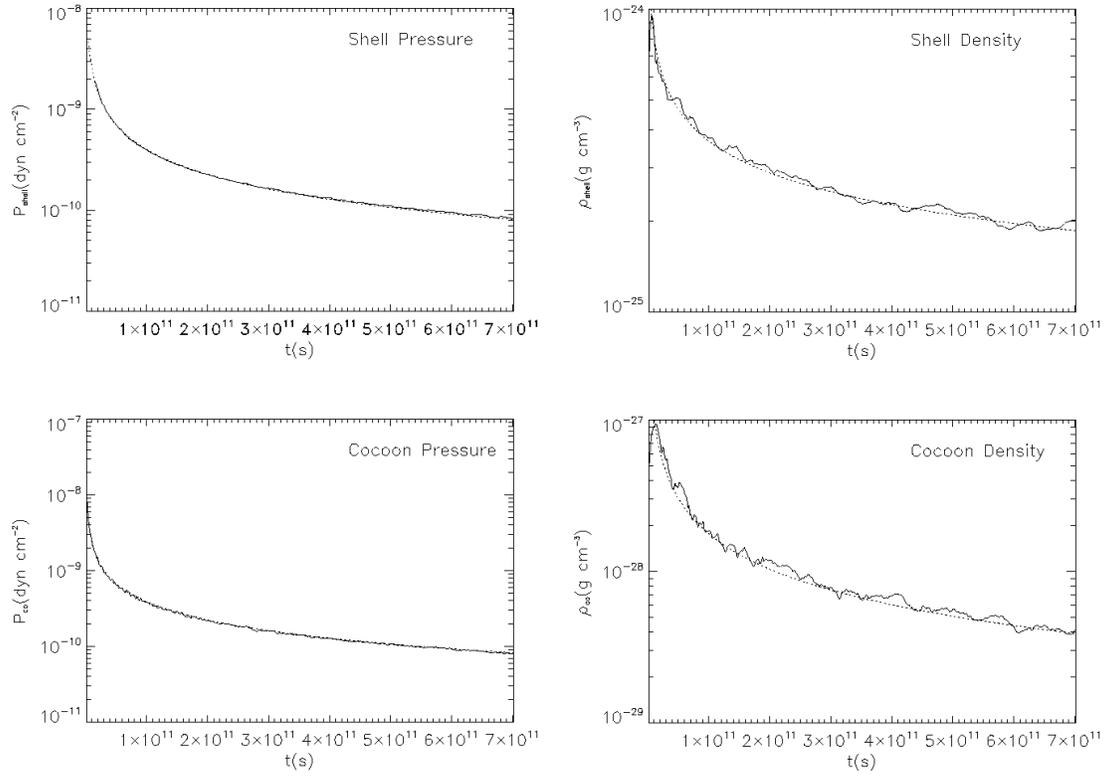}

      \caption{Pressure (left) and mass density (right) evolution in the shell (top) and cocoon (bottom) regions as a function of time. A fitting of the results is also shown for both variables (dotted line). This fit serves to estimate the simulation values at longer times. This extrapolation is strictly valid only if an homogeneous external medium and a constant injection energy rate are assumed. }

         \label{pressions_i_densitats}

   \end{figure*}

\end{appendix}



\begin{thebibliography}{}



\bibitem[1998]{aharonian98} Aharonian, F. A. \& Atoyan, A. M. 1998, NewAR, 42, 579
\bibitem[1974]{blandford1974} Blandford, R. D., \& Rees, M. J. 1974, MNRAS, 169, 395
\bibitem[2000]{Blundell00} Blundell, K. M.; Rawlings, S. 2000, AJ, 119, 1111
\bibitem[1970]{blum70} Blumenthal, G. R., \& Gould, R. J. 1970, Rev. Mod. Phys., 42, 237
\bibitem[2005]{bosch05} Bosch-Ramon, V., Aharonian, F.~A., \& Paredes, J.~M. 2005, A\&A, 432, 609
\bibitem[2005]{corbel05} Corbel, S., Kaaret, P., Fender, R.~P., et al. 2004, ApJ, 617, 1272
\bibitem[2002]{corbel02} Corbel, S., \& Fender, R.~P. 2002, ApJ, 573, L35
\bibitem[1983]{Drury83} Drury, L. 1983, SSRv, 36, 57
\bibitem[1991]{Falle1991} Falle, S. A. E. G. 1991, MNRAS, 250, 851
\bibitem[1974]{Faranoff1974} Fanaroff, B. L., \& Riley, J. M. 1974, MNRAS, 167P, 31F
\bibitem[2001]{Fender2001}Fender, R. P. 2001, MNRAS, 322, 31
\bibitem[2004]{Fender2004a} Fender, R. P. 2004, in Lewin W. H. G., van der Klis M., eds, Compact Stellar X-ray Sources. (Cambridge Univ. Press)
\bibitem[2005]{gallo05} Gallo, E., Fender, R.~P., Kaiser, C., Russell, D., et al. 2005, Nature, 436, 819
\bibitem[2003]{Gallo2003}Gallo, E., Fender, R.~P., \& Pooley, G.~G. 2003, MNRAS, 344, 60
\bibitem[2003]{heindl03} Heindl, W.~A., Tomsick, J.~A., Wijnands, R., \& Smith, D.~M. 2003, ApJ, 588, L97
\bibitem[2002]{Heinz02} Heinz, S. 2002, A\&A, 388, L40
\bibitem[2002]{heinz02b} Heinz, S. \& Sunyaev, R. 2002, A\&A, 390, 751
\bibitem[2002]{heinz06} Heinz, S. 2006, ApJ, 636, 316
\bibitem[1984]{Hillas1984} Hillas, A. M. 1984, ARA\&A, 22, 425
\bibitem[1997]{Kaiser97} Kaiser, C. R., \& Alexander, P. 1997, MNRAS, 286, 215
\bibitem[2004]{Kaiser04} Kaiser, C. R., Gunn, K. F., Brocksopp, C., \& Sokoloski, J. L. 2004, ApJ, 612, 332
\bibitem[1987]{Landau87} Landau, L. D., \& Lifshitz, E. M. 1987, Fluid Mechanics, 2nd English Edition (Oxford: Pergamon)
\bibitem[1989]{Leahy89} Leahy, J. P., Muxlow, T. W. B., \& Stephens, P. W. 1989, MNRAS, 239, 401
\bibitem[1996]{marti96} Mart\'{\i}, J., Rodriguez, L. F., Mirabel, I. F., \& Paredes, J. M. 1996, A\&A, 306, 449
\bibitem[1997]{mart97} Mart\'{\i}, J.M., M\"uller, E., Font, J.A., Ib\'a\~nez, J.M., \& and Marquina, A. 1997, ApJ, 479, 151
\bibitem[1992]{mirabel92} Mirabel, F., Rodriguez, L. F., Cordier, B., Paul, J., \& Lebrun, F. 1992, Nature, 358, 215
\bibitem[1999]{Mirabel99} Mirabel, I. F., \& Rodr\'iguez, L. F. 1999, ARA\&A, 37, 409
\bibitem[2007a]{paredes07} Paredes, J. M., Rib\'o, M., \& Bosch-Ramon, V., et al. 2007a, ApJ, 664, L39
\bibitem[2007]{Paredes07} Paredes, J. M., Mart\'{\i}, J., Ishwara Chandra, C. H., \& Bosch-Ramon, V. 2007b, ApJ, 654, L135
\bibitem[2005]{Paredes2005} Paredes J. M., 2005, in High Energy Gamma-Ray Astronomy: 2nd International
 Symposium, eds. F.~A. Aharonian, H.~J. V\"olk, \& D. Horns. AIP Conference Proceedings, 745, 93
\bibitem[2005]{pe+05} Perucho, M., Mart\'{\i}, J.~M., \& Hanasz, M. 2005, A\&A, 443, 863
\bibitem[2007]{Perucho07} Perucho, M. \& Mart\'i, J. M. 2007, MNRAS 382, 526 
\bibitem[2008]{Perucho08} Perucho, M. \& Bosch-Ramon, V. 2008, A\&A, 482, 917 
\bibitem[1999]{Protheroe1999}Protheroe, R. J. 1999, ADP-AT-98-9 [astro-ph/9812055]
\bibitem[2005]{Ribo2005} Rib\'o, M. 2005, ASPC, 340, 269
\bibitem[1980]{Rudnick1988} Rudnick, L., 1988, ApJ, 325, 189
\bibitem[2007]{Russell07} Russell, D. M., Fender, R. P., Gallo, E., \& Kaiser, C. R. MNRAS, 376, 1341
\bibitem[1974]{Scheuer1974} Scheuer, P. A. G. 1974, MNRAS, 166, 513
\bibitem[1959]{Sedov1959} Sedov, L. I. 1959, Similarity and Dimensional Methods in Mechanics (New York: Academic Press)
\bibitem[2008]{Sikora2008} Krzysztof Nalewajko, Marek Sikora, MNRAS, in press, [astro-ph/08103912]
\bibitem[2006]{tudose06} Tudose, V., Fender, R. P., \& Kaiser, C. R., et al. 2006, MNRAS, 372, 417
\bibitem[2000]{velazquez00} Vel\'azquez, P. F. \& Raga, A. C. 2000, A\&A, 362, 780  
\bibitem[1980]{zealey80} Zealey, W. J., Dopita, M. A., \& Malin, D. F. 1980, MNRAS, 192, 731


  

\end{thebibliography}
\end{document}